\documentclass{aa}  
\usepackage{graphicx}
\usepackage{txfonts}
\begin{document}
   \title{Geomagnetic storm dependence on the solar flare class}

   \author{Yu. I. Yermolaev 
          \and
          M. Yu. Yermolaev
          }
   \offprints{Yu. I. Yermolaev} 

   \institute{Space Research Institute, Russian Academy of Sciences, 
              Profsoyuznaya 84/32, Moscow 117997, Russia \\
              \email{yermol@iki.rssi.ru} 
              }
\date{Received January 23, 2006; accepted March 16, 1997} 

 
  \abstract
   {Solar flares are often used as precursors of geomagnetic storms. 
    In particular, Howard and Tappin (2005) recently published in A\&A 
    a dependence between X-ray class of solar flares and Ap and Dst 
    indexes of geomagnetic storms which contradicts to early published 
    results. 
   }
   {We compare published results on flare-storm dependences and discuss 
    possible sources of the discrepancy. 
   }
   {We analyze following sources of difference: (1) different intervals 
    of observations, (2) different statistics and (3) different methods 
    of event identification and comparison. 
   }
   {Our analysis shows that 
    magnitude of geomagnetic storms is likely to be independent on X-ray  
    class of solar flares. 
   }
   {}


   \maketitle
%

\section{Introduction} 

One of the important aims of solar-terrestrial physics is investigation of 
possible causes of geomagnetic storms on the Sun and in the interplanetary 
space. Storms are primarily generated by large, long-duration southward 
component of interplanetary magnetic field (IMF) (Burton et al. 
\cite{Burtonetal75}, Lyatsky and Tan 
\cite{LyatskyTan03}, Zhang et al.
\cite{Zhangetal06}) associated with interplanetary 
coronal mass ejections (ICME - magnetic clouds and ejecta) and corotating 
interaction regions (CIR) (see recent papers and reviews by Gopalswamy et 
al.
\cite{Gopalswamyetal05}, Kane 
\cite{Kane05}, Meloni et al.
\cite{Melonietal05}, Schwenn et al.
\cite{Schwennetal05}, Yermolaev et al.
\cite{Yermolaevetal05}, Yermolaev and Yermolaev
\cite{YermolaevYermolaev06}, and references therein).  

Solar flares were one of the first strong disturbances discovered on the Sun 
and they were considered as the important source of almost all interplanetary 
and geomagnetic disturbances during long time. Later, in the beginning of 
1970s, other powerful solar processes such as coronal mass ejections (CMEs) 
were discovered, and after the landmark paper by Gosling 
(\cite{Gosling93}) the situation 
has significantly changed, and now CME is considered almost as the unique 
cause of all interplanetary and geomagnetic disturbances (see recent reviews 
by Schwenn et  al.
\cite{Schwennetal05}, Yermolaev et al.
\cite{Yermolaevetal05} and references therein). 
Nevertheless the solar flares are often considered as a precursor of solar 
activity and used for prediction of interplanetary and geomagnetic 
disturbances (see recent papers by Park et al.
\cite{Parketal02}, Yermolaev et al.
\cite{Yermolaevetal05} and references therein).

Recently a statistic study of interplanetary shocks and accompanying events 
on the Sun and in the magnetosphere for 1998-2004 was published by Howard 
and Tappin 
(\cite{HowardTappin05}). 
In particularly there is Fig.7 showing a dependence between 
class of solar flares (X-ray measurements on GOES satellites) and value of 
geomagnetic storms (Ap and Dst indexes) with statistics of 103 pairs of 
events. On the basic of these data the authors indicated "a tendency for 
large flares to be associated with very large storms". This is very strong 
contention because if it would be true the class of solar flares could be 
used not only to predict an occurrence of magnetic storm but  also to  
predict a magnitude of it. Unfortunately authors (Howard and Tappin 
\cite{HowardTappin05}) 
did not compare this result with results of other papers. So, the aim of 
this paper is to compare this result with another published results on this 
topic. 

\section{Observations}

Published results on flare-storm dependence are presented in the table. 
Shrivastava and Singh 
(\cite{ShrivastavaSingh02}) and Howard and Tappin 
(\cite{HowardTappin05}) initially 
selected CME-magnetosphere pairs of events and then analyzed relation  
between classes (respectively, optic class in 1st paper and X-ray class 
in 2nd paper) of accompanying flares and magnetospheric disturbances (Ap 
index in 1st paper and Ap and Dst indexes in 2nd paper). Howard and Tappin 
(\cite{HowardTappin05}) 
addinionally selected events accompanied by interplanetary shocks. Although 
correlations between optic and X-ray classes of flares and between various 
geomagnetic indexes are sufficient low (Yermolaev and Yermolaev 
\cite{YermolaevYermolaev03b}), 
these papers say in favor of existence dependence between flare class 
and storm magnitude.   

%
\begin{table*}
\caption{Published results on correlation between solar flare class 
         and magnetosphere disturbance}             
\label{table:1}      
\centering                          
\begin{tabular}{c c l c c c l}        
\hline\hline                 
N & Statistics & Solar events & Magnetosphere & Time & Relation & Reference \\ 
  &            &              & events        & intervals &     & \\
\hline                        
   1 & 144     & Optic flare $>1$ (F, N, B) + CME & Ap & 1988-1993 & Yes &
Shrivastava \& Singh, \cite{ShrivastavaSingh02} \\      
   2 & 325     & X-ray flare $\ge M5$              & Dst& 1976-2000 & No &
Yermolaev \& Yermolaev, \cite{YermolaevYermolaev02a} \\
   3 & 325     & X-ray flare $\ge M5$              & Dst& 1976-2000 & No &
Yermolaev \& Yermolaev, \cite{YermolaevYermolaev03a} \\
     & 70      & X-ray flare $\ge M0 + SPE$        &    &      & No & \\
   4 & 103(?)  & X-ray flare $>$ C0 + CME + Shock& Ap, Dst&1998-2004 & Yes & 
Howard \& Tappin, \cite{HowardTappin05} \\
\\ 
\hline                                   
\end{tabular}
\end{table*}

Similar analysis of solar, interplanetary and magnetospheric events for 
1976-2000 had been published by Yermolaev and Yermolaev 
(\cite{YermolaevYermolaev03a}) 
(see also preliminary publication by Yermolaev and Yermolaev 
\cite{YermolaevYermolaev02a}) 
where the same dependence had been presented (see Fig.5 in paper by 
Yermolaev and Yermolaev 
\cite{YermolaevYermolaev03a}). The dependence of magnitude of 
325 storms on X-ray class ($ \ge M5$) of solar flares was presented 
on top panel 
of the figure 5 in the paper and the same dependence for 70 flares 
($\ge M0$) accompanied by Solar Particle Events (SPEs) - on bottom panel. 
In two panels the data have been selected with (1) location of solar 
flare on solar disc - west (open symbols) and east (closed), and 
(2) time delayed between flare and corresponding  storm - 2 - 4 days 
(high probability of event relation, triangles), 1.5-2 and 4-5 days 
(intermediate probability, rhombs), and 1-1.5 and 5-6 days 
(low probability, circles). No tendency of increase in value of storms 
with increasing class of solar flares was observed.

Thus two different results were obtained in different studies. 
Possible causes of this difference will be discussed in next section 
of paper. 

\section{Discussion} 

Two papers, which indicate the existence of flare-storm relation 
(Shrivastava and Singh 
\cite{ShrivastavaSingh02} and Howard and Tappin 
\cite{HowardTappin05}), have common 
feature in method of data selection (This feature is absent in papers 
by Yermolaev and Yermolaev
\cite{YermolaevYermolaev02a}, 
\cite{YermolaevYermolaev03a}): initial selection of 
CME-magnetosphere pairs of events and consequent analyses of relation 
between classes of accompanying flare and magnetospheric disturbances.  
So, necessary condition of existence of flare-storm relation is likely 
to be existence of CME-storm relation. This condition is not clearly 
stated in papers by Shrivastava and Singh 
(\cite{ShrivastavaSingh02}) and Howard and Tappin 
(\cite{HowardTappin05}) and this hypothesis requires further investigations. 

It is difficult to compare results of papers by Shrivastava and Singh 
(\cite{ShrivastavaSingh02}) and Yermolaev and Yermolaev 
(\cite{YermolaevYermolaev02a}, 
\cite{YermolaevYermolaev03a}) because they were 
obtained with use of absolutely different methods of event definition 
and classification. Nevertheless, several considerations, which will 
be applied below to comparison of results in papers by Howard and 
Tappin 
(\cite{HowardTappin05}) and Yermolaev and Yermolaev 
(\cite{YermolaevYermolaev02a}, 
\cite{YermolaevYermolaev03a}), may be of 
interest in future data analyses.   

In addition to mentioned above methodical difference (initial 
selection of CME-magnetosphere pairs of events) in papers by 
Shrivastava and Singh 
(\cite{ShrivastavaSingh02}) and Yermolaev and Yermolaev, 
(\cite{YermolaevYermolaev02a}, 
\cite{YermolaevYermolaev03a})
 there are three main possible causes of discrepancy: 
(1) different intervals of analysis, (2) different statistics and 
(3) different methods of event identification and comparison. 
Yermolaev and Yermolaev 
(\cite{YermolaevYermolaev03a}) studied 25-year interval (more 
than 2 solar cycles from 1976 up to 2000) while Howard and Tappin 
(\cite{HowardTappin05}) 
investigated only 7-year interval near maximum of 23-rd 
solar cycle (1998-2004). As well known, the magnetic storms are 
generated by different types of solar wind disturbances (magnetic 
clouds, MC, or corotating interaction regions, CIR, which are 
generated by CME or fast streams from coronal hole, respectively)  
during different phases of solar cycle (see, for instance, Fig. 6 
in paper by Yermolaev and Yermolaev 
\cite{YermolaevYermolaev02b}). It is possible to 
suggest that averaging data over solar cycle could mask indicated 
dependence but this hypothesis requires further investigations. 
 
The higher statistics in paper by Yermolaev and Yermolaev 
(\cite{YermolaevYermolaev03a}) 
indicate in favour of absence of storm dependence on class of flare. 
For instance, extremely strong geomagnetic storm on March, 1989 
(Dst = -589 nT) can be associated with large (but not extremely large) 
flares with class X1-X5 and this event does not agree with suggested 
flare-storm relation. 

As has been shown (Yermolaev et al., 
\cite{Yermolaevetal05}, Yermolaev and Yermolaev
\cite{YermolaevYermolaev03a}, \cite{YermolaevYermolaev06}) 
result of comparison of different events on the Sun, 
in the interplanetary space and in the geomagnetosphere strongly 
depends on methods of event identification and comparison procedures. 
Unfortunately methodical problems related to dependence under study 
are very schematically discussed in paper by Howard and Tappin 
(\cite{HowardTappin05}) 
and it makes impossible to search for cause of result discrepancy
in features of methods. 

Available data allow us to discuss only problem of selection of flares 
with various classes for comparison with magnetic storms. Howard and 
Tappin 
(\cite{HowardTappin05}) included C-class flares in the analysis. As well known 
CMEs (not flares) generated interplanetary disturbances and then magnetic 
storms (Gosling
\cite{Gosling93}), and flares can be used only as indicator of 
solar activity which can result in CMEs and interplanetary disturbances. 
On the other hand, association flares and CMEs decreases with decreasing 
class of flares (Kahler et al.
\cite{Kahleretal89}). In recent paper by Yashiro et al. 
(\cite{Yashiroetal05}) 15\% and 30\% flare-CME associations 
were obtained respectively for 
disc and limb flares  with class range of C3-M1. So, C-class flares, 
included by Howard and Tappin 
(\cite{HowardTappin05}) in analysis, could not improve 
correlation between class of flares and Dst index during magnetic storms. 

\section{Conclusions}

Thus, our analysis of published results allows one to make preliminary 
conclusions. 

   \begin{enumerate}
      \item There is no any correlation between X-ray class of solar flares 
            and magnitude of corresponding geomagnetic storms. 
      \item If one selects initially CMEs and corresponding geomagnetic storm 
            and then solar flares accompanying CMEs, for solar flares obtained 
            by this way a slight positive correlation between these parameters 
            is likely to be observed. 
   \end{enumerate}

Such a correlation would be very important for space weather prediction and 
reliability of it requests further investigations.

\begin{acknowledgements} 
       The authors are grateful to all databases for data used in the analysis. 
       Work was in part supported by RFBR, grant 04-02-16131.  
\end{acknowledgements}

\end{document}